# Can Scientific Journals be Classified in terms of Aggregated Journal-Journal Citation Relations using the *Journal Citation Reports*?

Loet Leydesdorff

School of Economics (HEC), Université de Lausanne, Switzerland

&

Amsterdam School of Communications Research (ASCoR), University of Amsterdam

Kloveniersburgwal 48, 1018 CE Amsterdam, The Netherlands

loet@leydesdorff.net ; http://www.leydesdorff.net

**Abstract**

The aggregated citation relations among journals included in the *Science Citation Index* provide us with a huge matrix which can be analyzed in various ways. Using principal component analysis or factor analysis, the factor scores can be used as indicators of the position of the *cited* journals in the *citing* dimensions of the database. Unrotated factor scores are exact, and the extraction of principal components can be made stepwise since the principal components are independent. Rotation may be needed for the designation, but in the rotated solution a model is assumed. This assumption can be legitimated on pragmatic or theoretical grounds. Since the resulting outcomes remain sensitive to the assumptions in the model, an unambiguous classification is no longer possible in this case. However, the factor-analytic solutions allow us to test classifications against the structures contained in the database. This will be demonstrated for the delineation of a set of biochemistry journals.

## 1. Introduction

The hierarchical classification of scientific disciplines and their corresponding literatures resounds with an ancient tradition of theology and philosophy, and it appeals to the reductionist program in the modern philosophy of science (Hempel & Oppenheim, 1948; Nagel, 1961). Hierarchical approaches consider the sciences as originating from a common source. In the Middle Ages theology in the center was assisted by philosophy as its servant (*ancilla*), but the scholastic organization tolerated the other sciences to be organized in a circle around this center. The organization of the sciences into a Trivium and a Quadrivium was later reproduced in the organization of faculties at the universities. Among the seven 'artes liberales,' however, a hierarchy was no longer assumed.

The advent of modernity turned the tables. Natural philosophy became identified with physics as the sole science with a foundation in mathematical principles (Kant, 1781, [2]1787; Newton, 1687). The vision of a unity of science was inspired by the idea that all sciences might eventually be reduced to the so-called 'scientific worldview' (Neurath *et al.*, 1929). Strong remnants of this unified vision of science are still present in the philosophy of science (Popper, [1935] 1959). However, the concept of a disunity of science has gained support both from the perspectives of the history and philosophy of science and from evolutionary perspectives on the development of the sciences (Dupré, 1993; Galison & Stump, 1995). From these perspectives, the sciences develop in terms of fields in which specialties span competitive research fronts (Hull, 2001). The specialties and their corresponding literatures can be expected to operate in parallel.

Derek de Solla Price (1963) conjectured that specialties would begin to exhibit 'speciation' when the carrying community grows larger than a hundred or so active scientists (Crane, 1972; Kochen, 1983). Furthermore, the proliferation of scientific journals can be expected to correlate with this proliferation of communities because new communities will wish to begin their own journals (Price, 1965). New journals are organized within existing frameworks, but the bifurcations and other network dynamics feed back on the historical organization to the extent that new fields of science and technology become established and existing ones thus reorganized (Van den Besselaar & Leydesdorff, 1996).

The evolutionary dynamics operates in the present and potentially reorganizes the historical dynamics from the perspective of hindsight. Classification schemes of the sciences have reflected these two perspectives along the time axis (Leydesdorff, 2002a). Narin (1976) first advocated the development of an *ex ante* classification scheme of scientific journals, since this would allow us to follow the development of the sciences given an analytic scheme for the evaluation (Narin *et al.*, 1972; Narin & Carpenter, 1975).

One disadvantage of such a fixed scheme is that new journals are either disregarded or organized within previously established categories.

Alternatively, one can backtrack from a most recent understanding of the organization of the journal literature (*ex post*) into origins that are currently still relevant, and consider the history as having provided only the variation. However, this assumes that one can provide a current understanding of the organization of the journal literature, for example, in terms of science policy priorities (Leydesdorff *et al.*, 1994). At another level of aggregation, the difference between the two perspectives on the time axis (cited versus citing) has also been discussed in terms of the difference between bibliographic coupling among cited documents (Marshakova, 1973) versus co-citation among documents from the perspective of the citing documents (Small & Griffith, 1974). The 'citing' dimension provides a selective focus in the present on the archive of science which can potentially be 'cited.'

## 2. The research question

The aggregated journal-journal citation reports provided by the *Journal Citation Reports* (*JCR*) of the *Science Citation Index* (*SCI*), in principle, contain the information for an understanding of the whole of the sciences in terms of disciplines and specialties, both in the cited and the citing dimensions. In this study, I shall investigate (i) whether such a top-down decomposition is feasible, (ii) which methods are required for the analysis and why, and (iii) how sensitive the consequent classification remains for the initial assumptions. The conclusion will be that the exact solution (using principal component analysis) does often not provide meaningful results, while the rotated component analysis does. However, the latter is based on assumptions, for example, about the number of relevant dimensions. The results of the top-down decomposition therefore remain dependent on initial choices. The consequences will be elaborated for the set of chemistry journals and in a next decomposition for the subset of journals indicating the specialty of biochemistry. Using this example, I will demonstrate how the analytical clarification allows us to test the quality of a logical classification against the structures contained in the database.

The ISI itself provides a classification of journals at the level of the database which has been based on intuitive criteria (Pudovkin & Garfield, 2002). Hitherto, the mere size of the database has been prohibitive for breaking it down without any assumptions about a hierarchy (Glänzel & Schubert, 2003). Alternatively, researchers have focused on local contexts of journal sets and the optimization of the delineation of specialties in a bottom-up approach (Doreian & Fararo, 1985; Leydesdorff, 1986, 1987; Tijssen *et al.*, 1987). For example, Leydesdorff & Cozzens (1993) developed an optimization procedure that stabilizes approximations of eigenvectors of the network as representations of specialties. Local optimizations have also been used for clustering document sets in terms of other indicators. For example, during the 1980s the ISI invested considerably in generating a *World Atlas of Science* using the alternative method of co-citation clustering (Garfield *et al.*, 1975; Small & Garfield, 1985; Leydesdorff, 1987).

The recent expansion of memory capacities to gigabytes makes the handling of large databases in a single pass nowadays feasible. In a previous paper, I experimented with loading the *Journal Citation Report* of the *Social Science Citation Index* 2001 as an SPSS systems file and then generating factor-analytical results about the commonalities in the variance in the citation patterns among these journals (Leydesdorff, 2004b). In this study, I extend this approach using the *Journal Citation Report* of the *Science Citation Index* 2003 containing the citation patterns of 5907 journals. The size of the database forced me to be reflexive on the choices in terms of similarity criteria and clustering algorithms. As I shall show, certain trade-offs between idealizations and pragmatic considerations still have to be made.

### *2.1 The primacy of the factor-analytic approach*

Social network analysis enables us to use graph theory for inferencing (e.g., Otte & Rousseau, 2002). The focus in graph theory, however, is on relations. For example, one can study the hyperlink structures at the Internet using tools commonly available nowadays in programs for social network analysis (like UCINET and Pajek). A network can be expected to contain a structure because some relations are more important than others. For example, citation relations among journals in solid-state physics are more dense than the relations between journals in this specialty and journals dedicated to plant physiology.

In a series of studies (Leydesdorff, 2004a, 2004b; Leydesdorff & Jin, 2005) I experimented with using bi-connected components as a graph-analytical approach to distinguishing robust clusters of journals. This technique was specifically developed for the purpose of distinguishing robust clusters (Knaster & Kuratowski, 1921; Moody & White, 2003). However, graph analysis remains focused on relations as different from structural positions. For example, one can no longer position the bi-connected components in relation to one another. Furthermore, relations provide the variation which changes over time, while structure can be expected to remain more stable in the background. The analysis of this latent structure, however, requires a structural approach, that is, an approach at the systems level (Burt, 1982; Leydesdorff, 1993b).

The matrix of citation relations among journals is overwhelmingly empty. Despite the low thresholds in the ISI database (when only single citations are to a large extent excluded), more than 97% of the cells in the matrix are empty (Leydesdorff & Jin, 2005). Thus, the citation relations among journals are heavily structured by virtually empty spaces among them. This strong structuration does not mean that journals occur only in dense clusters which are otherwise not related. Some stars in the network may have a wide span, but the expectation remains that these ties are weak (Granovetter, 1973). If clustering occurs, it is likely to be local given the otherwise overwhelmingly empty space.

Structure selects upon the variation. While variation in the relations can exhibit randomness, selection operates deterministically. The multidimensional space is first spanned by the citation patterns of the journals as vectors. A (re)description of this variation in terms of the selecting structures is then possible using principal component or factor analysis. This procedure decomposes the space resulting from the aggregation of the vectors in an analytical and parsimonious way. Unlike graph analysis, factor analysis is not based on analyzing the *relations* among the journals that span the space as variation, but functional to the understanding of the *positions* that journals obtain within the structure as a consequence of the relational aggregation. In other words, by using factor analysis one can study the dimensions of the selecting (in this case, citing) system.

Given this systems perspective, the delineation of the system of reference becomes crucial. Like other forms of multivariate analysis, factor-analytic models are sensitive to the inclusion or exclusion of single variables. Although individual journals may be marginal to a field, they can have very pronounced citation patterns and this individual pattern may have a significant impact on the aggregate of citation relations. The journals with high loadings on a factor representing a specialty are often highly specialized and therefore not major journals. Major journals can span across specialties and may therefore exhibit interfactorial complexity (Leydesdorff & Jin, 2005; Pudovkin & Garfield, 2002; Van den Besselaar & Heimeriks, 2001). In terms of the vector-space model, this means that they are positioned in-between the main axes that span the system as a multi-dimensional space.

### *2.2 Rotation and the number of factors*

While the variation in the citations is operationally organized into structure by the selections in the citing articles, the resulting structure can be expected to contain a next-order variance of main dimensions or so-called eigenvectors. Principal component analysis allows us to rewrite the total variance in a matrix in terms of these eigenvectors. This operation is strictly analytical: a principal component is nothing more than a linear combination of the vectors so that the largest eigenvector can be extracted. This eigenvector can then be attributed an eigenvalue. The eigenvalue for a given factor measures the variance in all the variables which is accounted for by that factor (Garson, 2004).

When the first eigenvector is extracted from the variance, the remaining variance can again be considered and a second eigenvector can be extracted using the same technique. The resulting components are thus completely independent of each other or—in other words—these axes are orthogonal. The extraction of more eigenvectors does not change the eigenvalue of previously extracted components. Sequential components will extract ever less variance and thus this rewrite enables us to achieve an economy in the representation. For example, one can ask for the first hundred components in the citation matrix of the *JCR*. If the analyst wishes more detail, more components can be extracted without affecting the previously obtained results.

Since the components are based on a linear combination of vectors, the main components may be spanned in directions where the multi-dimensional space happens to be empty. Two or more major clusters of points may incidentally add up in a single direction without any of them occupying the resulting position. The designation of the component can in this case pose a problem for the analyst. The noted emptiness of the matrix makes it unlikely that there will always be a one-to-one correspondence between densities in the multi-dimensional space and the analytical dimensions of this space.

In order to achieve this maximum correspondence between the densities in the multidimensional space and the main components, the components can be rotated (either under the restriction of orthogonality among the axes or not). However, the criterion for this optimization is no longer to explain as much *variance* as possible in the data, but to find common factors in the set that explain the *covariance* (or correlation) between the variables. While both methods (principal component analysis and rotated component analysis) allow for data reduction, the rotated component matrix is no longer an analytical rewrite of the original data, but it is necessarily based on an assumption about the number of factors that span the multidimensional space. In other words, one needs a model for the explanation of correlations, while principal component analysis is still part of the descriptive statistics (Kim, 1978).

In the case of the *Journal Citation Reports* one has no *a priori* reason to assume that any given number of factors (e.g., one hundred) will provide us with an optimal representation of the structure in the database. This condition suggests the use of principal component analysis for the decomposition. As noted, the designation of the components may remain problematic in this case because the unrotated components are based on aggregations of clusters. The factor-analytic classification schemes based on rotation, however, add our assumptions to the classification. These assumptions can be informed by the principal component analysis or be based on other (e.g., pragmatic) criteria.

For example, one may wish to ask how the journal set can be decomposed into twelve main categories (Glänzel & Schubert, 2003). Given this research question, a rotated factor matrix would be most informative. However, the rotated solutions based on twelve or thirteen components can be different. In practice, these differences may be small so that optimization techniques can successfully be applied.

### *2.3 Factor scores and factor loadings*

The factor loadings in a rotated component matrix are (by definition) equal to the correlation ($r$) of the hypothesized dimension with the variable. As noted, small and specialized journals may represent a highly specific citation pattern to a larger extent than larger journals. However, one can additionally compute (and save) the factor scores for each journal as a case. The factor score is the value attributed to this unit on the dimension of the eigenvector as a constructed variable.[1]

[1] The factor scores are computed by multiplying the factor score coefficients with the standardized values of the variables. The factor score coefficient matrix is computed from the factor pattern matrix or, in the case of orthogonal rotation, from the factor structure matrix (Kim, 1975, at pp. 487 ff.)

In this study, we are interested precisely in the factor scores because they provide us with a measure for the contribution of the *cited* journal to each specific factor in the multidimensional space of *citing* journals. For the reasons specified above, the factor scores remain in the rotated case dependent on the number of factors extracted. In the unrotated case, however, the factor scores are exact, but the meaning of the axes for the classification may be more difficult. These unrotated component scores provide us with precise values of the contribution of journals to the citation patterns in the database.

In summary, the theoretical expectation is that some small and highly specialized journals may have a very high factor *loading* after rotation, indicating specificity. In the unrotated case, the larger journals may be more pronounced. In both cases, the factor *scores* of the small journals are likely to be smaller than the ones of the larger journals on their respective dimensions. Thus, there may be a trade-off between the interpretability of the results (after rotation) versus the replicability and analytical precision of the results (before rotation). Let us study the need and the consequences of this trade-off empirically in the case of the *Journal Citation Report* for the year 2003.

## 3. Methods and materials

The *Journal Citation Report* of the *Science Citation Index* 2003 covers 5907 journals; 5714 of these journals are processed not only from the cited side, but also 'citing.' This means that the ISI not only registers whether a journal is cited by using all the citations in the citing journals, but the citing journals are among the set which is actively scanned for citations when composing the database (Garfield, 1979; Leydesdorff, 1993a; Wouters, 1999).

This data contains 971,502 unique citation relations among journals on a total of over 3 * $10^7$ (= 5714 x 5907) possible citation relations. This is equal to 2.88%.[2] The number of citations included is 17,604,594 or an average of 18.12 citations/journal-journal relation. However, all these distributions are heavily skewed, that is, ranging from a within-journal citation rate of 49,870 for the *Journal of Biological Chemistry* to 277,838 times only two citations per journal-journal relation.[3]

The asymmetrical matrix of 5714 citing journal patterns as the variables and 5907 cited journals as the cases provides the starting point for our analysis. The asymmetry in the matrix warrants that we do not violate the assumption of factor analysis that the number of cases should be larger than the number of variables.[4] As specified above, our research question focuses on the position of the cited pattern of journals within the multi-dimensional space spanned by the citing patterns at the systems level. In other words, the matrix is not squared and also asymmetrical.

---

[2] If one substracts the diagonal cells for self-citations (5714), the remainder is 871,502 and 15,703,356, respectively. However, one should keep in mind that at the level of journals apparent 'self-citations' include also citations among articles from different authors, but within the same journal. Thus, they can be considered as 'within-journal' citations. I decided to keep these values in. In order to correct for the effect of self-citations, one may consider normalizing the values on the main diagonal using one of the algorithms which have been proposed for this purpose, for example, by Price (1981) and Noma (1982).

[3] In most cases the value of one is reset to zero in the database, but 145 times the value of one was retained in this set (for unknown reasons).

[4] The number of degrees of freedom in a matrix is the minimum number of variables or cases minus one. Thus, if the number of cases is relatively small, the number of factors meaningfully to be extracted can be constrained. Factor analysis, therefore, is not recommended for the reduction of square matrices (cf. Klavans & Boyak, manuscript, at p. 3).

Unfortunately, the current versions of SPSS (11 and 12) have a systems limitation for the workspace of 2,097,151 kbytes or, equivalently, somewhat more than 2 Gbytes. Although this allows for reading and storing a systems file with 5714 variables, it is not sufficient for the factor extraction. In several runs, we were able to manage matrices with appr. 3600 variables, but there is an additional limitation of showing only 3000 points in visualizations. This latter limitation can be circumvented using other programs for the visualization. Excel and Pajek will be used for the visualizations below.

Since the problem is purely technical and one can expect it to be resolved in one of the next versions of SPSS (or perhaps using other operating systems), I used the 3174 citation pattern of citing journals with a threshold of a variance ≥ 8.0. This threshold is somewhat arbitrary, but the variance is the decisive parameter contributing to the factor structure (Leydesdorff & Bensman, in preparation). Journals with low variance in the citation patterns will tend to be relatively small. After more than 3000 journals are included, the further addition of smaller journals cannot be expected to influence the structure of the multi-dimensional vector-space significantly.[5]

[5] The systems file was also imported into programs for social network analysis like Ucinet and Pajek. Although these programs have been developed primarily for graph analysis, Ucinet contains a routine for the factor analysis. This routine seems unlimited in terms of the capacity, but in practice we did not obtain results from it when feeding it with this size of matrices after several days of computation.

## 4. Results

### *4.1 The extraction of eigenvectors (descriptive statistics)*

In the multi-dimensional space of 3174 vectors, 416 eigenvectors had a value larger than one. The screeplot of the first hundred of these eigenvectors is exhibited in Figure 1. The distribution suggests that there is a bit of discontinuity after seven factors. Eleven factors explain more than one percent of the variance in the matrix. In order to allow for a comparison with the journal categories specified by Glänzel & Schubert (2003)—in a follow-up study—I decided to focus on the twelve main components as indicators of fields, and then to compare these with the hundred first components as potentially more specialized indicators.

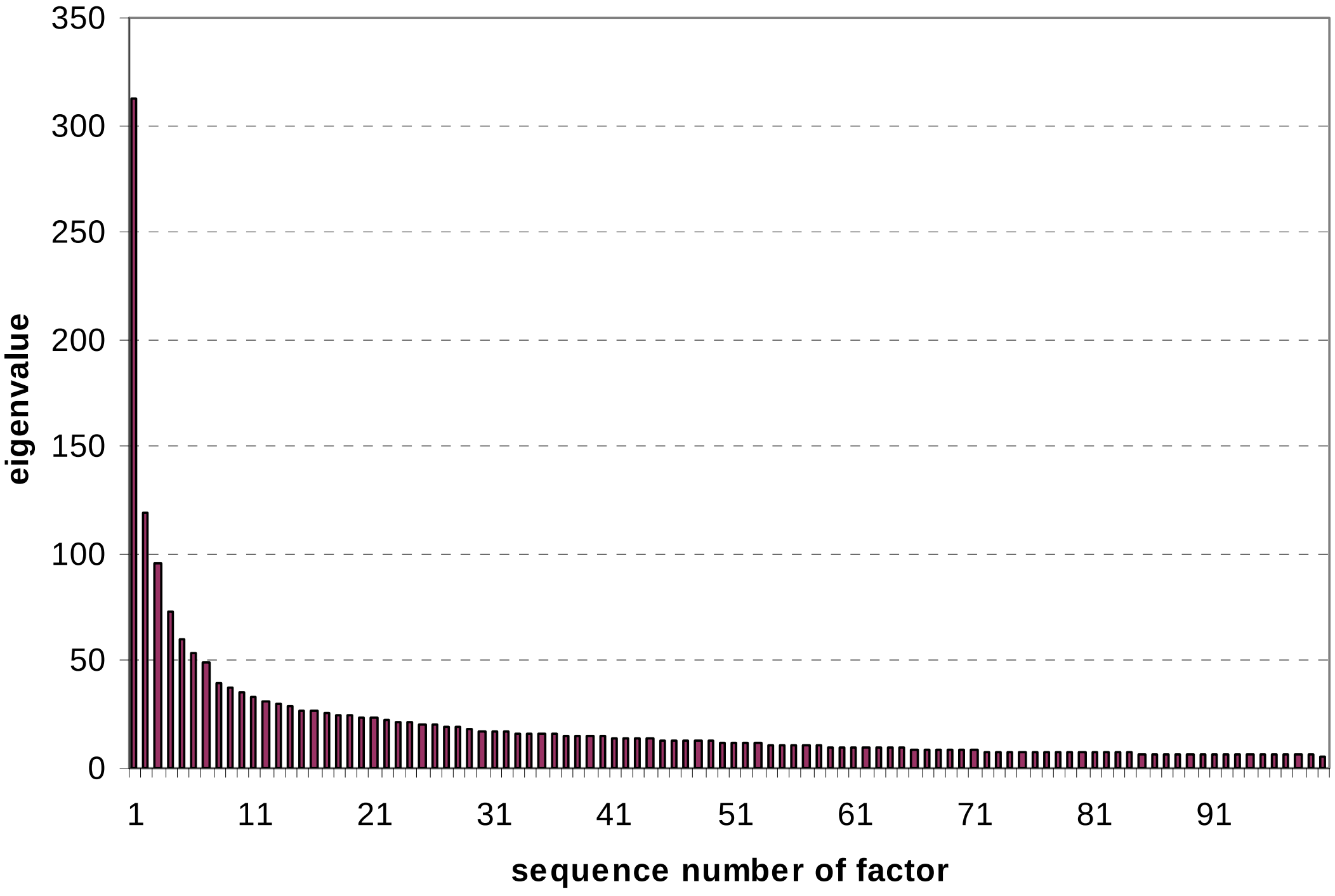

**Figure 1**. Screeplot of the first 100 eigenvectors.

*4.2 Principal component analysis*

Figure 2 exhibits the plot of the factor scores for the two major components of the database using the 15 cited journals (that is, cases) which have a value on either of these scores larger than 10.0 or smaller than –10.0. (The journals associated with factor score values between –10 and +10 are suppressed in order to keep the legends in the charts readable.)

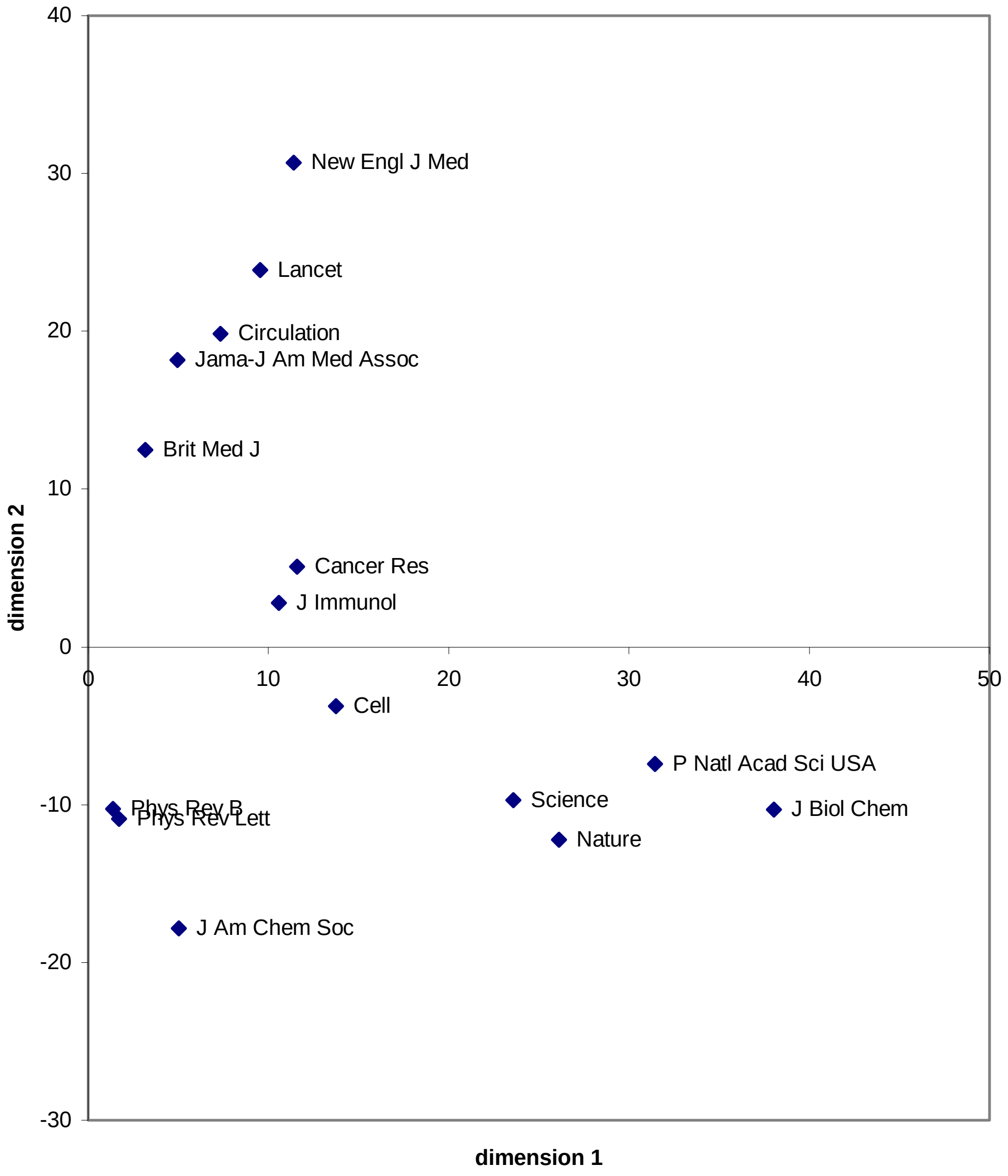

**Figure 2**. Factor scores on the first and second principal component of 15 journals with absolute values of their respective factor scores larger than ten (on both factors).

The first dimension is (not surprisingly) spanned by the large effect of citations of major journals in the life sciences (the *Journal of Bioliogical Chemistry* and the *PNAS*). It is noteworthy that in this 'general science' dimension no journals have a factor score on the negative side < –1.0.  As one can conclude from inspection of Figure 3, however, only a few hundred (major) journals show a coupling on this dimension, while the large majority of the journals (more than 4000) exhibit a slightly negative tendency (between zero and –1) on this dimension. These specialist journals are organized in dimensions mainly orthogonal to the general science journals.

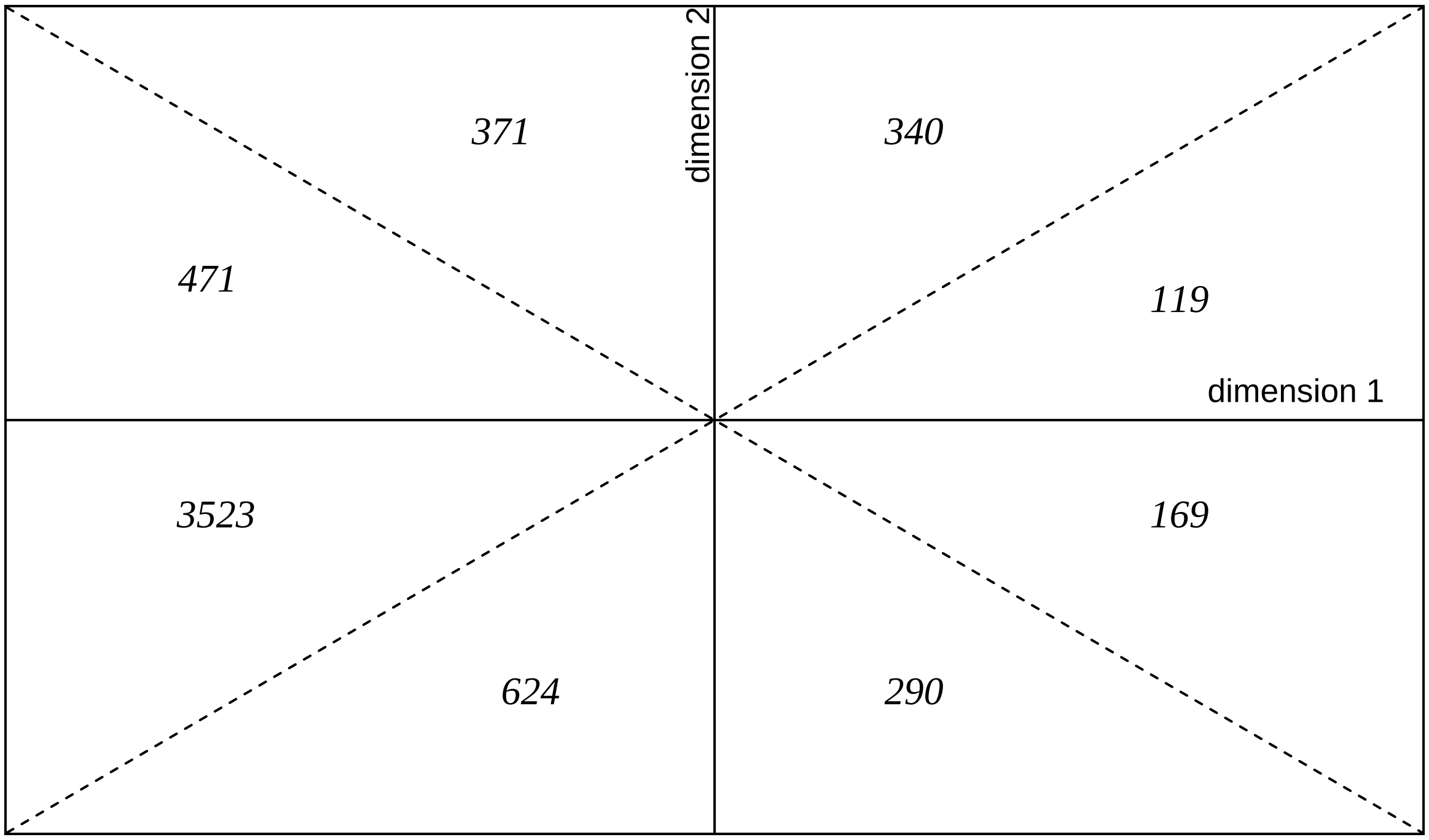

**Figure 3**. Distribution of the journals according to their factor scores on the two main dimensions of the database.

The second dimension is spanned between the medical axis on the positive side and the natural sciences on the negative side. These arrows are championed by the *New England*

*Journal of Medicine* and the the *Journal of the American Chemical Society*, respectively. *Science* and *Nature* mainly gain their status as interdisciplinary journals by exhibiting leading factor scores on both the major journals in the life sciences (dimension 1) and on the negative side of the second axis representing the natural sciences.

In summary, these results make it possible to divide the database into major components by using different queries. For example, the 290 journals in the lower half of the second quadrant of Figure 3 can be expected to be leading journals in the natural sciences which are not at the same time general science journals. The top half of the figure (quadrants I and IV) can be considered as representing the more than one thousand journals which focus on medicine from a clinical perspective (as against a bio-medical one).

The further extraction of more components is illustrated in Figure 4 by exhibiting the factor scores of 13 journals on the third and the fourth dimensions with thresholds similar as in Figure 2 (that is, factor scores larger than 10 or smaller than –10; the figure is log-scaled for reasons of presentation).[6] In this graph, the sciences are further decomposed, first in a dimension between the life sciences and the natural sciences (dimension 3), and vertically in a fourth dimension between physics and chemistry.[7] As expected, the life sciences are more related to chemistry than to physics along this vertical dimension. However, previous to this split a distinction between the *Journal of Biological Chemistry*

[6] Bensman (2004) argued for the logarithmic transformation of citation data on theoretical grounds. We intend to discuss this issue in a separate study (Leydesdorff & Bensman, in preparation).
[7] One more journal (the *Lancet*) has factor scores between -10 and +10 on one of the two factors, but factors scores between -1 and +1 on the other. Because the logarithm is negative between zero and one, the *Lancet* was not included in this visual representation. However, its inclusion would not have made a difference for the interpretation.

and the *Journal of the American Chemical Society* prevails as a third dimension of the database.

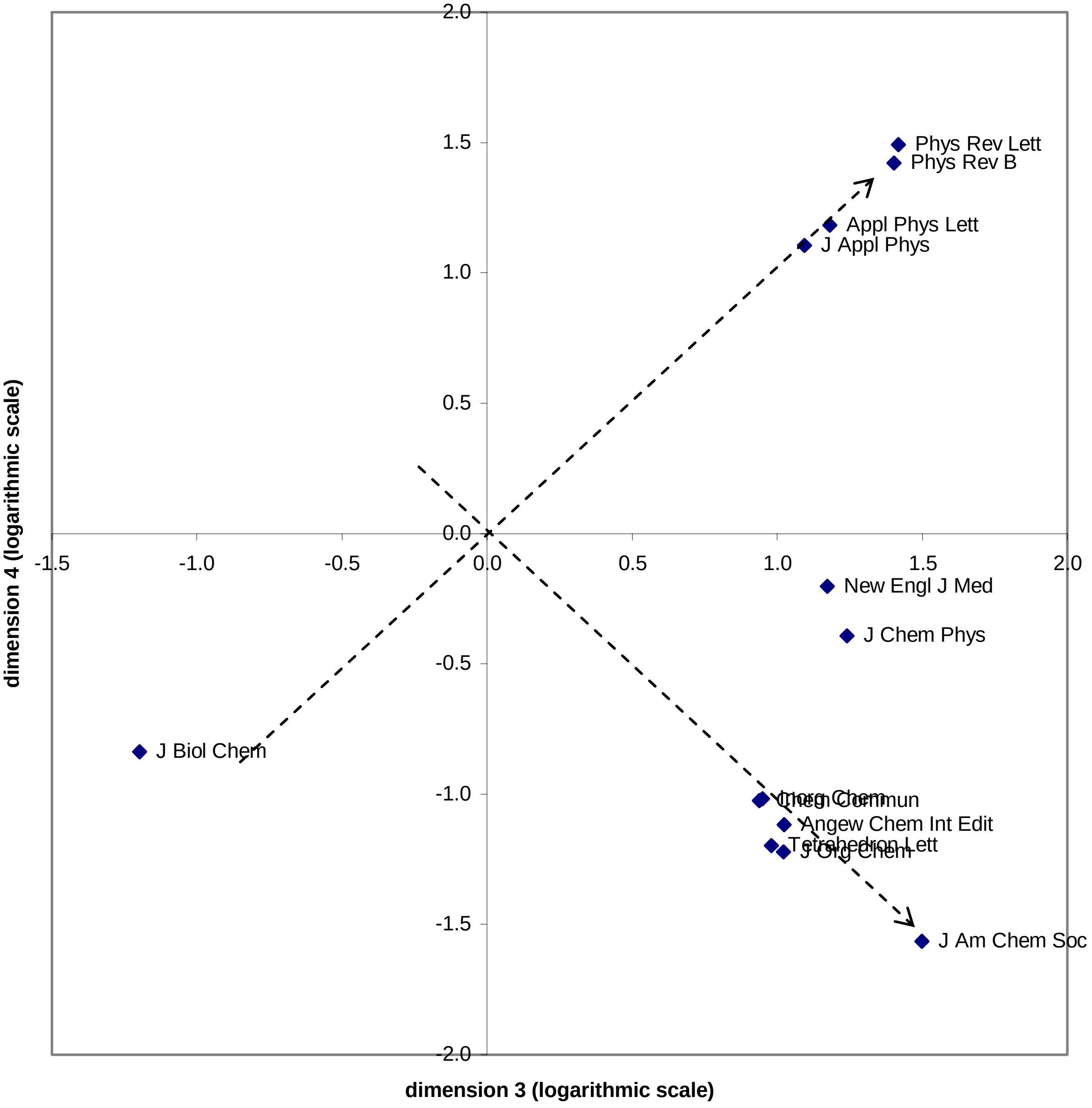

**Figure 4.** Distribution of 13 journals with unrotated factor scores larger than ten or smaller than minus ten on the third and fourth dimensions of the database (log scaled).

Figure 4 suggests that a better representation of the division between physics and chemistry would be obtained upon rotation. This is indicated by the arrows added to the figure, and will be pursued in a next step on the basis of the assumption of twelve major factors for reasons specified above.

### *4.3 Rotated component analysis*

Inspection of the rotated component matrix teaches us that in this model (based on twelve factors) 'chemistry' retains the third position, but physics follows only as the tenth dimension. This means that the density of the cloud of physics journals in the multi-dimensional space is lower than in chemistry. The order of the rotated factors informs us about the covariation or correlation, while the principal component analysis informs us about the contribution to the variance. As one can read from Table 1, which summarizes the results of the rotated component analysis, major physics journals also play a role in the fifth dimension. However, this fifth component is more on the applied side since it also includes journals in engineering.

| Factor designation | Highest factor loading (n = 3174) | Highest factor score (n = 5907) | Nr of journals with a factor score > 0 |
|---|---|---|---|
| 1. Molecular Biol. | *Biochem Biophys Res Co* | *J Biol Chem* | 774 |
| 2. Medicine | *Presse Med* | *New Engl J Med* | 922 |
| 3. Chemistry | *Chin J Chem* | *J Am Chem Soc* | 585 |
| 4. Cancer Research | *Cancer* | *Cancer Res* | 767 |
| 5. Physics & Eng. | *Mat Sci Eng B-Solid* | *Phys Rev B* | 650 |
| 6. Neurosciences | *Brain Res Rev* | *J Neurosci* | 976 |
| 7. Immunology | *Immunol Cell Biol* | *J Immunol* | 1065 |
| 8. Circulation | *Coronary Art Dis* | *Circulation* | 1375 |
| 9. Ecology | *Ecol Lett* | *Ecology* | 908 |
| 10. Physics | *Phys Rep* | *Phys Rev Lett* | 651 |
| 11. Geosciences | *Earth Planet Sci Lt* | *J Geophys Res* | 597 |
| 12. Neurology | *J Neurol Neurosur PS* | *Neurology* | 1107 |

**Table 1**. Rotated component model in twelve factors for the citation patterns of 3174 journals citing 5907 journals included in the database.

The comparison of the previous Figure 4 with Figure 5 below demonstrates the advantages of the rotation: upon rotation the axes can be designated unambiguously. (Because of the low values on the other axis, the pictures based on rotated solutions can no longer be log-scaled. Note the interfactorial position of the *Journal of Chemical Physics* in the first quadrant.) The major axes of the rotated solution are relatively robust when more factors are added, but the classification is now necessarily based on a model. The assumption about the number of factors, however, could not be warranted in terms of the descriptive statistics because of the continuity in the screeplot (Figure 1).

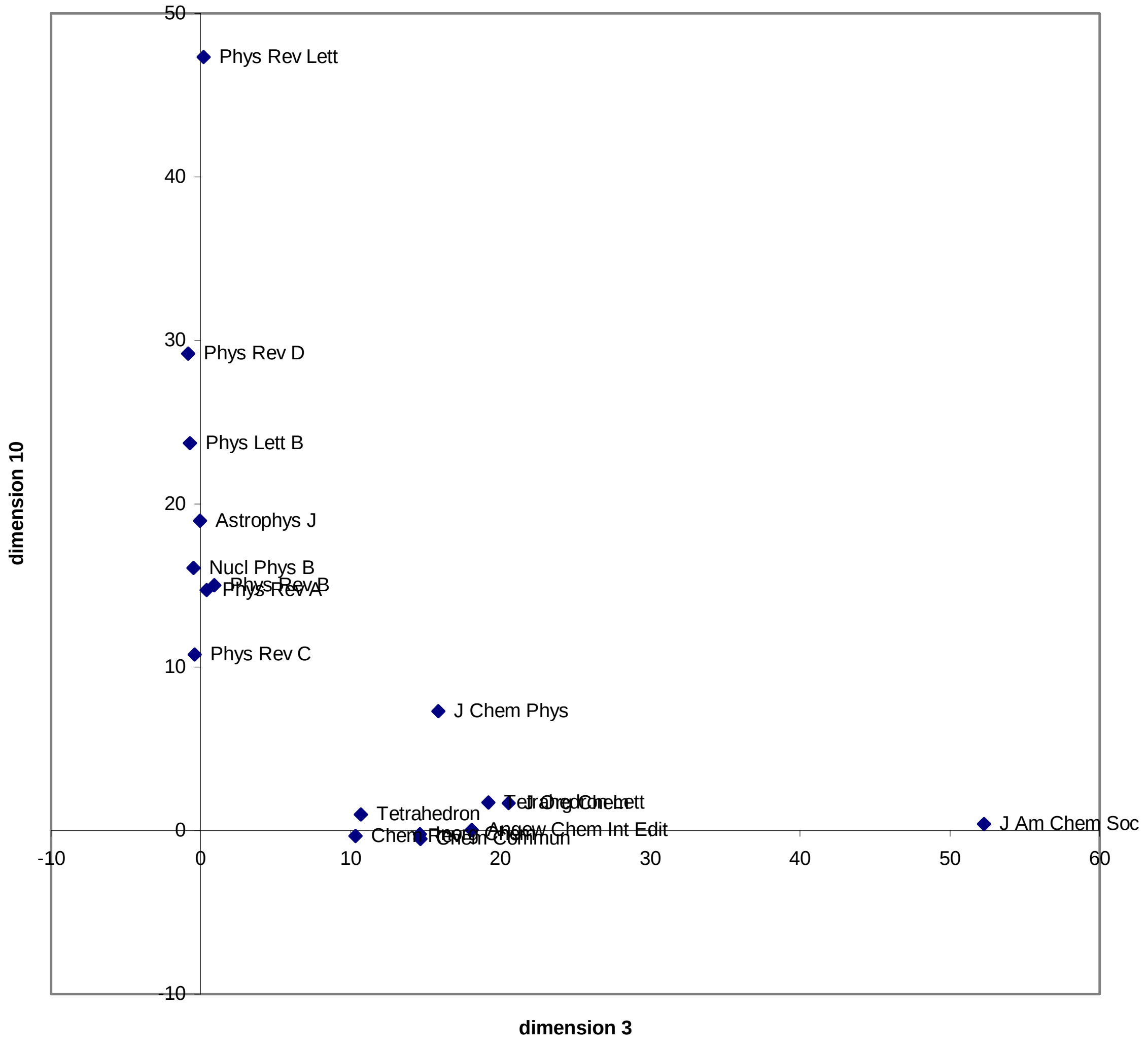

**Figure 5**. Distribution of 17 journals with factor scores larger than 10 or smaller than –10 on both the third or the tenth dimension of the rotated component analysis.

For example, the exact factor scores for the unrotated 99th and 100th principal components are plotted in Figure 6. This picture teaches us, among other things, that there is an axis in the variance between the *IEEE Transactions on Communications* and the *IEEE Transactions on Microwave Theory and Techniques*. This density can further be investigated, for example, by choosing a journal which is relevant locally in the two respective citation environments (Figure 8); but let us first consider the rotated solution for the journals along this axis at the systems level (Figure 7).

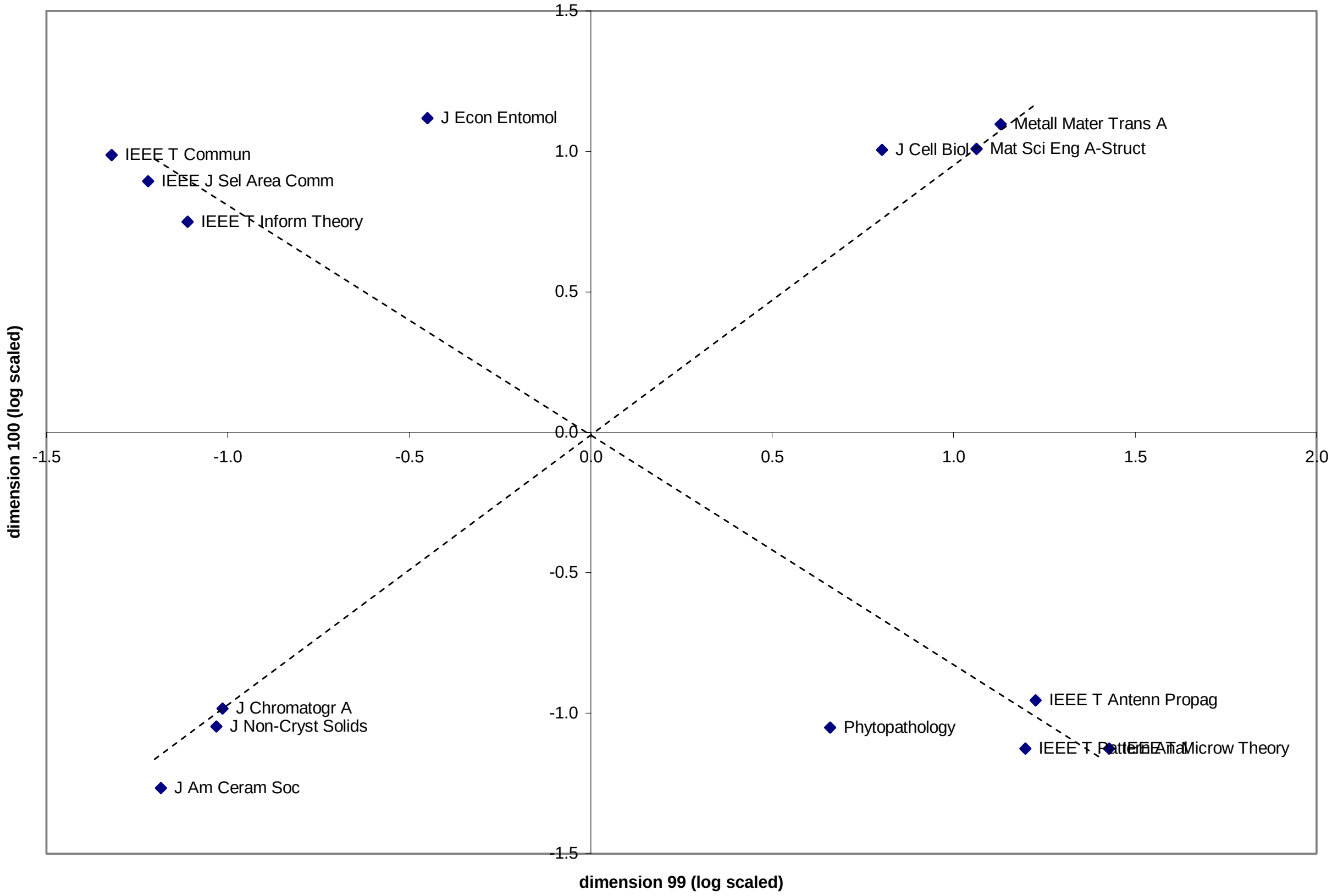

**Figure 6**. Distribution of 14 journals with factor scores larger than 10 or smaller than –10 on the 99th and 100th dimension of the unrotated component analysis (log scaled).

The dimensions with highest factor loadings and factor scores in the rotated solutions happen to be the 97th and the 98th in a model of extracting one hundred factors after rotation. These two dimensions are plotted against each other in Figure 7. Note that if the model allows for sufficient dimensions, negative factor scores remain relatively low (Leydesdorff, 2004b). This means that each dimension in the database is nearly orthogonal or relatively self-contained. In other words, the dimensions are not defined in opposition to one another, but independently, that is, using angles of ninety degrees.

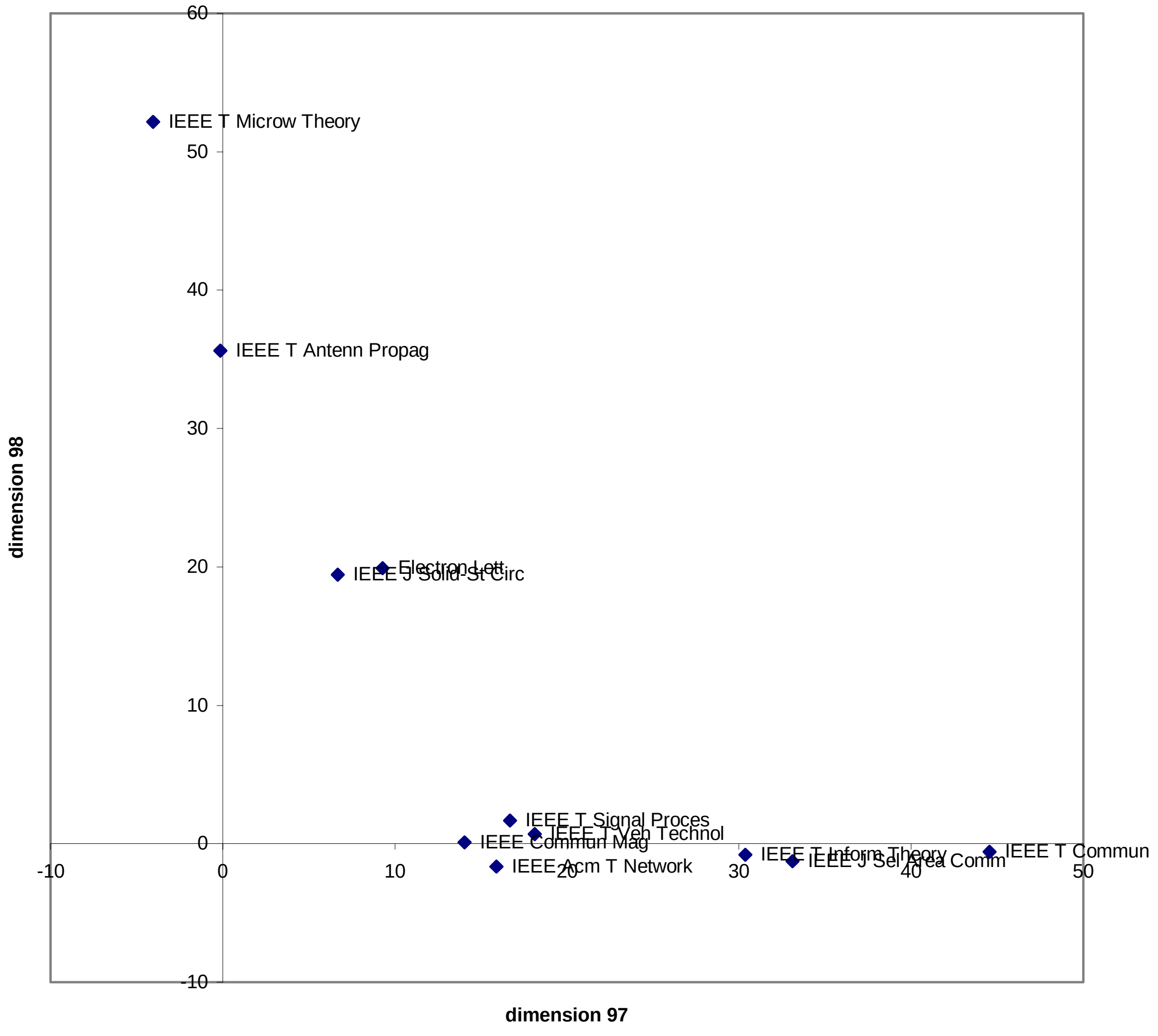

**Figure 7**. Rotated factor scores of 11 journals with absolute values of factor scores > 10 on dimensions 97 and 98 in the rotated solution of 100 eigenvectors.

By focusing on an intermediate journal with loadings on both components, like *Electronics Letters,* one can also extract this local environment using our previously developed methods (Leydesdorff, 1986; Leydesdorff & Cozzens, 1993). Thirty-nine journals are included in the citation environment at a threshold of 0.5% of the total citations from or references to this journal. Figure 8 shows that both specialties function on the margins of the field of opto-electronics (which can itself be considered as a high-tech specialty of applied physics). The fine structure of this field is further revealed by the factor analysis provided in Table 2 and accordingly penciled into Figure 8.

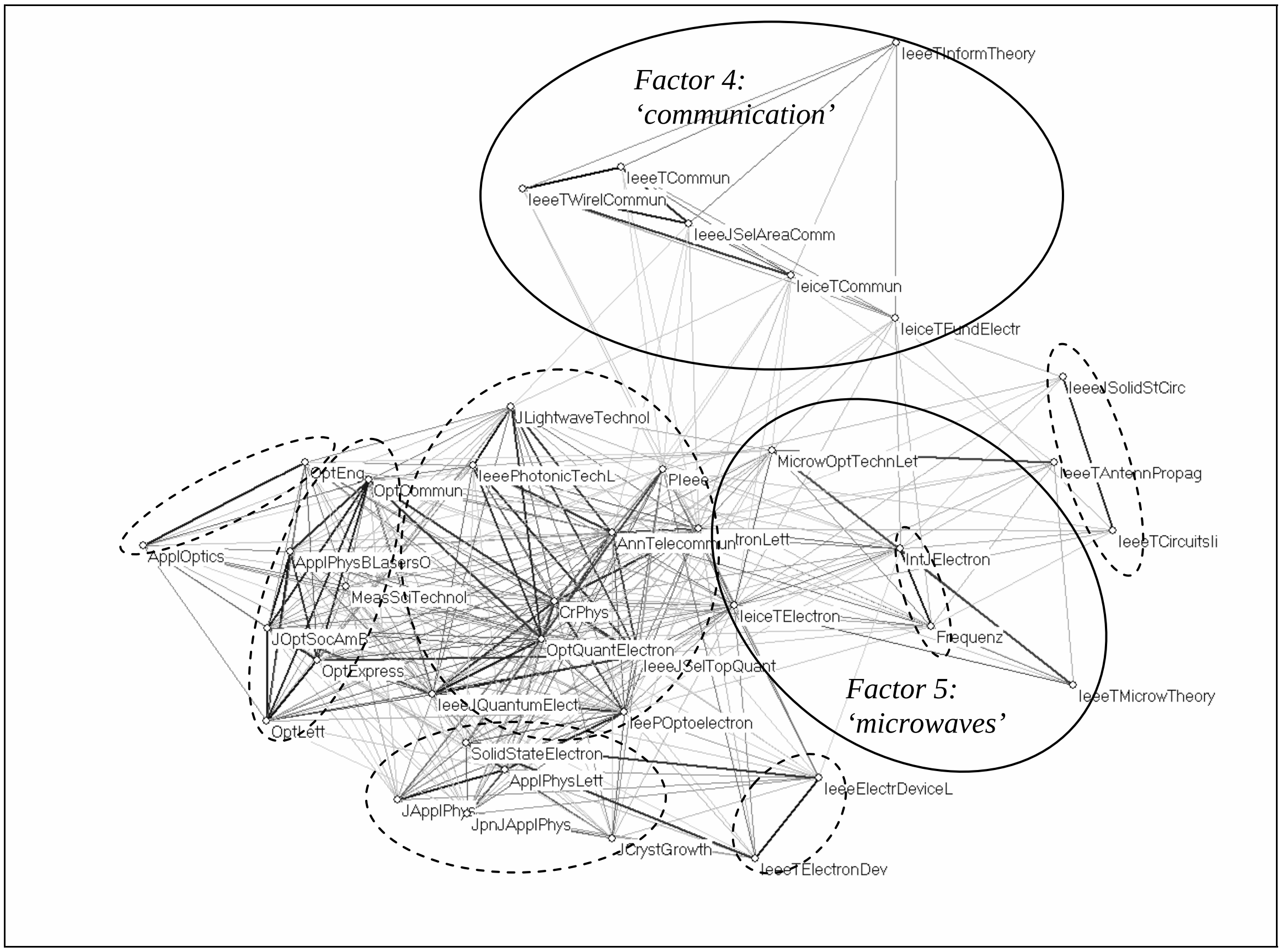

**Figure 8**. Thirty-nine journals in the citation environment of the *Electronics Letters* (cosine ≥ 0.2).

**Rotated Component Matrix** [a]

| | Component | | | | | | | | |
|---|---|---|---|---|---|---|---|---|---|
| | 1 | 2 | 3 | 4 | 5 | 6 | 7 | 8 | 9 |
| IEEE PHOTONIC TECH L | .945 | | .101 | | | | | | |
| J LIGHTWAVE TECHNOL | .923 | | | | | | | .127 | |
| OPT QUANT ELECTRON | .834 | .155 | .386 | -.124 | | | | .145 | |
| ANN TELECOMMUN | .796 | .142 | .225 | | .132 | | .437 | | |
| IEEE J SEL TOP QUANT | .772 | .108 | .490 | -.100 | | | | .210 | |
| CR PHYS | .720 | .479 | .325 | | | | .280 | | |
| ELECTRON LETT | .690 | .170 | | .172 | .209 | .126 | .507 | | .156 |
| IEEE J QUANTUM ELECT | .607 | .332 | .581 | -.102 | | | | | |
| APPL PHYS LETT | .132 | .931 | .145 | | | .146 | | | |
| J APPL PHYS | | .880 | | | | .177 | | | |
| JPN J APPL PHYS | | .814 | | | | .150 | | | |
| J CRYST GROWTH | | .797 | | -.100 | | | | | |
| IEE P-OPTOELECTRON | .578 | .762 | .103 | | | | | | |
| SOLID STATE ELECTRON | | .695 | | | | .661 | | | |
| P IEEE | | .657 | | .162 | | .438 | | | .266 |
| J OPT SOC AM B | .181 | .103 | .934 | | | | | | |
| APPL PHYS B-LASERS O | | .121 | .919 | -.104 | | | | .109 | |
| OPT LETT | .214 | | .906 | | | | | .210 | |
| OPT EXPRESS | .297 | | .822 | | | | | .356 | |
| OPT COMMUN | .338 | | .802 | | | | | .367 | |
| IEEE T WIREL COMMUN | | | | .940 | | | | | |
| IEEE T COMMUN | | | | .937 | | | | | |
| IEEE J SEL AREA COMM | | | | .907 | | | | | |
| IEICE T COMMUN | | | -.142 | .778 | .182 | | -.124 | | |
| IEEE T INFORM THEORY | | | | .662 | -.136 | | .105 | | |
| IEICE T FUND ELECTR | | | -.114 | .530 | | | .287 | | .328 |
| MICROW OPT TECHN LET | .200 | | | | .905 | | .268 | | |
| IEEE T ANTENN PROPAG | | | | | .784 | -.130 | | | |
| IEEE T MICROW THEORY | | | | | .766 | .121 | .338 | | |
| IEICE T ELECTRON | .253 | .319 | | | .620 | .350 | .284 | | .277 |
| IEEE T ELECTRON DEV | | .212 | | | | .941 | | | |
| IEEE ELECTR DEVICE L | | .274 | | | | .906 | | | |
| FREQUENZ | .166 | | -.106 | | .317 | | .883 | | |
| INT J ELECTRON | .117 | | | | .319 | | .864 | | .152 |
| OPT ENG | .276 | | .246 | | | | | .864 | |
| APPL OPTICS | | | .355 | | | | | .864 | |
| MEAS SCI TECHNOL | | .237 | .191 | -.111 | | | .106 | .629 | |
| IEEE J SOLID-ST CIRC | | | | | | | | | .965 |
| IEEE T CIRCUITS-II | | | | | | | .152 | | .932 |

Extraction Method: Principal Component Analysis.
Rotation Method: Varimax with Kaiser Normalization.

a. Rotation converged in 7 iterations.

**Table 2**. Rotated component matrix for the citation environment of *Electronics Letters* (threshold: 0.5% of the total cited or total citing of the journal).

### *4.4 Hierarchical decomposition of the factorial groupings*

Table 1 above provided a summary of the factor designation of the twelve-dimensional model after rotation. Figures 4 and 5 showed that the order of the rotated factors can be very different from that of the principal components. The criterion of the ordering is different. For example, at the fourth place of the rotated solution a factor focussing on cancer research organizes the database more strongly than the two physics factors (five and ten). Furthermore, the various biomedical sets exhibit overlap, but the covariation within this group can be organized in different directions. More fine-grained delineations—as between the 'neurosciences' (factor 6) and 'neurology' (factor 12)—became visible even within these first twelve dimensions.

Each of the dimensions can be characterized by those journals which have a positive factor score on this dimension, because journals with a negative score on a dimension definitely do not belong to the group thus indicated. Let us now proceed with the further decomposition of the smallest among these groupings, that is, the factor designated as 'chemistry' containing 585 journals. Thus, we extract the set of 585 journals with a positive factor score on the (third) dimension. As noted, this dimension is indicated most strongly by the *Journal of the American Society of Chemistry (JACS).* One can consider this set as the universe of journals relevant for 'chemistry,' and then proceed with the further analysis of this smaller set.

Note that this set of 'chemistry' journals is delineated from the set of the 'life science' journals which is spearheaded by the *Journal of Biological Chemistry (JBC)*. *JBC* and *JACS* entertain negative factor scores on each other's dimension. I shall use this independence in the dimensionality below to study how the definition of a set of 'biochemistry' journals is affected by choosing one perspective or the other.

Of the 585 journals thus defined, 15 journals are not included in the database in terms of their citing behaviour. Thus, a matrix of 570 variables versus 585 cases can be generated. The extraction of eigenvectors provides 88 factors with an eigenvalue larger than unity, and ten factors have an eigenvalue larger than 10. For reasons of consistency in the decomposition, however, let us assume again a twelve-factor model.

Figures 9 and 10 visualize the vector space of 542 of these journals without and with the labels attached, respectively. (The other 28 journals are no longer related to the set at the threshold level of cosine $\geq$ 0.2 and were therefore removed from the visualizations of the vector space.) These pictures show the strong structure in the set, which is further specified in Table 3 on the basis of the rotated factor matrix, strictly analogously to Table 1 above.

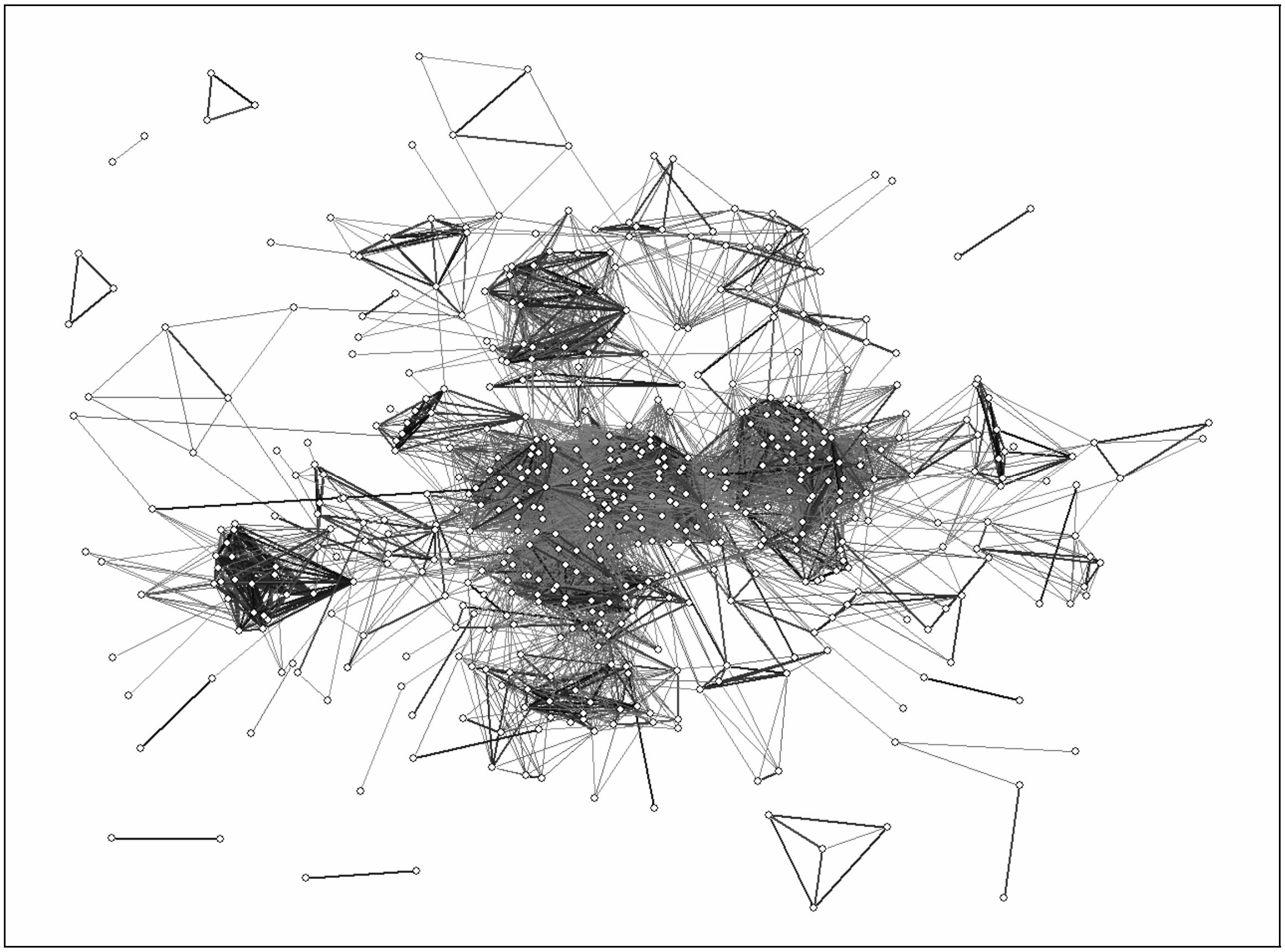

**Figure 9**. Unlabeled representation of the vector space (cosine ≥ 0.5) of the citation patterns of 542 chemistry journals with factor scores on the 'chemistry' dimension > 0.

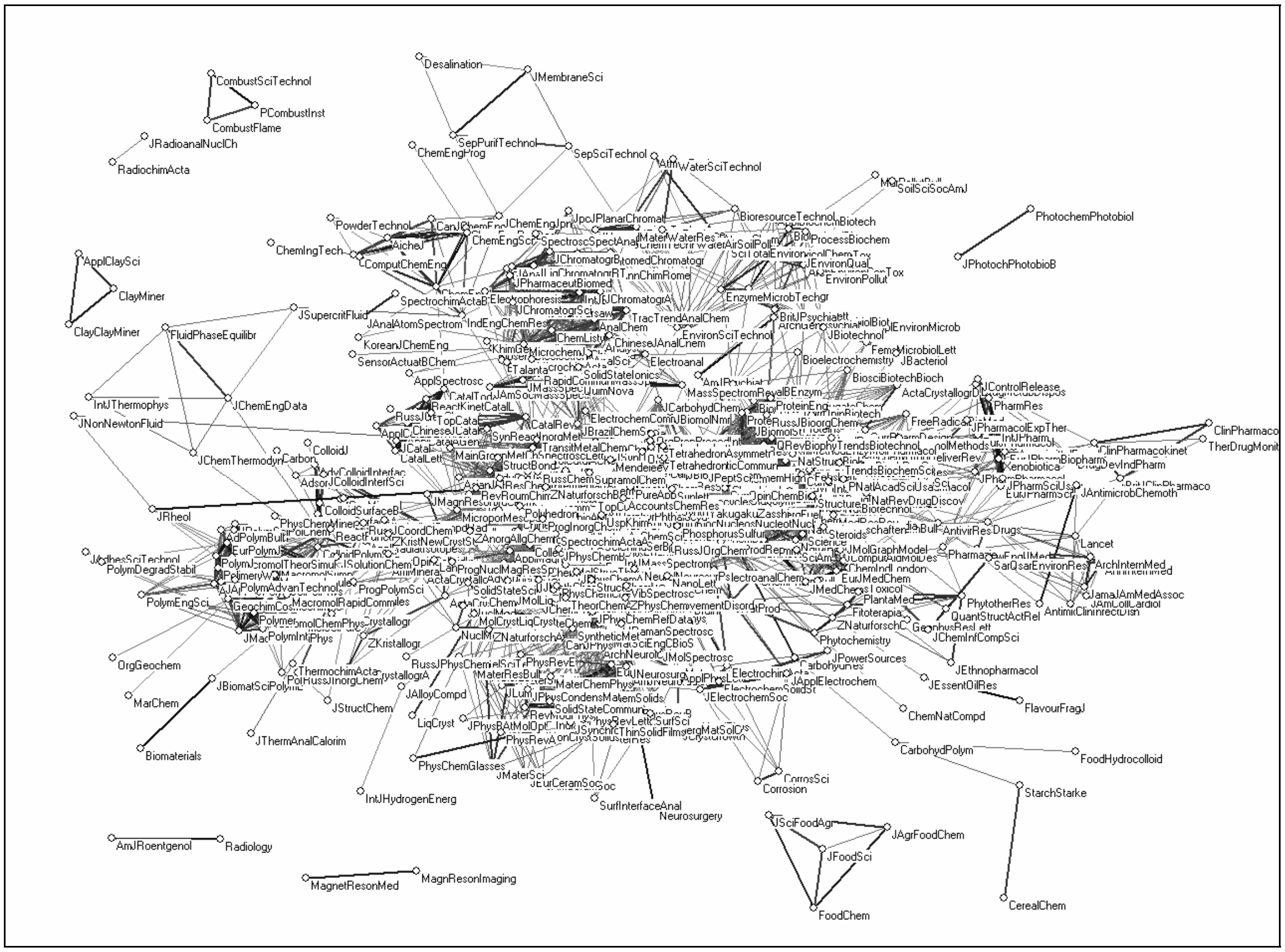

**Figure 10**. Labeled representation of vector space (cosine ≥ 0.5) of the citation patterns of 542 chemistry journals with factor scores on the 'chemistry' dimension > 0.

| Factor designation | Highest factor loading (n = 570) | Highest factor score (n = 585) | Nr of journals with a factor score > 0 |
|---|---|---|---|
| 1. Org. Chemistry | *Tetrahedron* | *Tetrahedron Lett* | 94 |
| 2. Biochemistry | *FEBS Lett* | *PNAS* | 97 |
| 3. Inorg. Chemistry | *Inorg Chem Comm* | *Inorg Chem* | 106 |
| 4. Chem Phys | *Theor Chem Acc* | *J Chem Phys* | 148 |
| 5. Analytical Chem. | *Combust Flame Anal* | *Anal Chem* | 94 |
| 6. Polymers | *Polym Int* | *Macromolecules* | 94 |
| 7. Solid State Phys | *Appl Phys-Mater* | *Phys Rev B* | 109 |
| 8. Chemical Eng. | *Chem Eng J* | *J Catal* | 115 |
| 9. Environm. Chem. | *Chemosphere* | *Environ Sci Technol* | 102 |
| 10. Electrochem. | *Electrochem Commun* | *J Electrochem Soc* | 196 |
| 11. Medical Chem. | *Drugs* | *New Engl J Med* | 133 |
| 12. Pharmacology | *J Pharm Pharmacol* | *Int J Pharm* | 136 |

**Table 3.** Twelve-factor model of the set of 570 journals citing 585 journals with positive factor scores on the 'chemistry' dimension.

The *size* of the specialties in terms of numbers of journals seems normally distributed, with a means at 118.7 and a standard deviation of 30.4. (In the previous case, the mean was 864.8 and standard deviation 240.5.) In the next section we repeat the decomposition for one lower level, and then we will find, as might be expected, the specialty structure of the discipline.

### *4.5 Decomposition at the specialty level*

Figure 11 shows the vector-space model of the 97 journals subsumed under the second factor extracted, which was designated in Table 3 as ‘biochemistry.’ The dense graph in the center of the figure confirms the previous delineation: these 97 journals can be considered the ‘biochemistry’ set with relevant citation environments in the domain of ‘chemistry.’ However, some journals may additionally be related to the set of journals indicated above as the ‘life sciences,’ and which were not included in the ‘chemistry’ set.

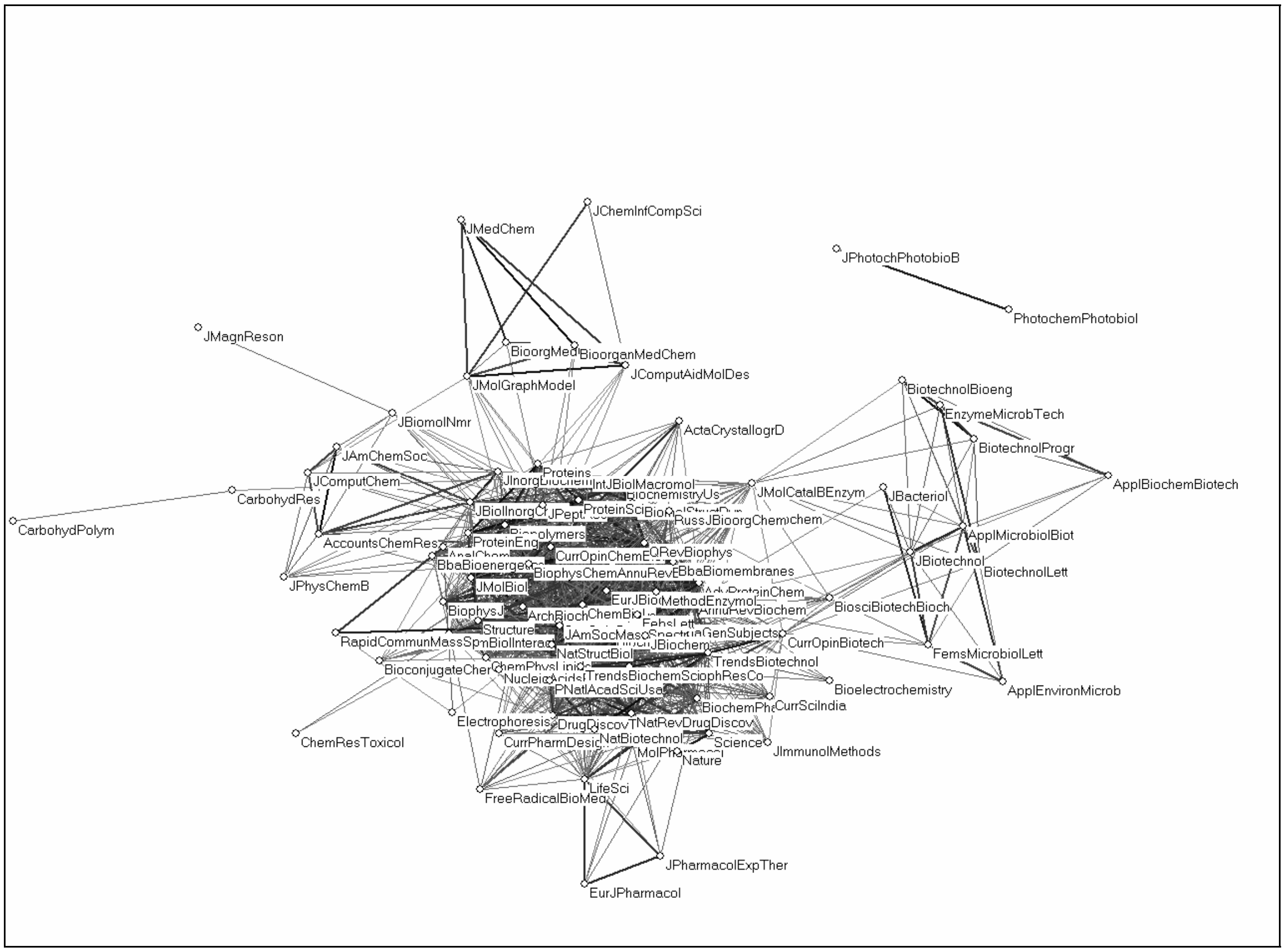

**Figure 11**. Ninety-one 'biochemistry' journals within the 'chemistry' set related at the level of cosine ≥ 0.5

| Factor designation | Highest factor loading (n = 96) | Highest factor score (n = 97) | Nr of journals with a factor score > 0 |
|---|---|---|---|
| 1. Enzymology | *Protein Sci* | *Biochemistry US* | 27 |
| 2. Life sciences | *Science* | *Nature* | 23 |
| 3. Biochemistry | *Chem-Biol Interaction* | *Biochem J* | 27 |
| 4. Chemistry | *J Am Chem Soc* | *J Am Chem Soc* | 25 |
| 5. Medical Chemistry | *J Med Chem* | *J Med Chem* | 19 |
| 6. Biotechnology | *Biotechnol Progr* | *Biotechnol Bioeng* | 26 |
| 7. Microbiology | *FEMS Microbiol Lett* | *J Bacteriol* | 17 |
| 8. Analytical Chem | *Anal Chem* | *Anal Chem* | 17 |
| 9. Pharmacology | *J Pharmacol Exp Ther* | *J Pharmacol Exp Ther* | 23 |
| 10. Photo | *J Photoch Photobio B* | *Photochem Photobiol* | 31 |
| 11. Carbohyd Res | *Carbohyd Polym* | *Carbohyd Res* | 23 |
| 12. Crystallography | *Acta Crystallogr A* | *Acta Crystallogr A* | 23 |

**Table 4**. Twelve-factor model of the set of 97 biochemistry journals with positive factor scores on the second dimension of the chemistry set of 570 journals.

Table 4 provides a summary of the rotated citation matrix in a format comparable to the previous ones. Note that the vector-space model (using the cosine among the vectors for the visualization) visualizes the unrotated dimensions, while the rotated factor model exhibits the fine structure of the specialty. The local densities can further be analyzed, as was done above for the citation environment of the *Electronic Letters* (Figure 8). The size of the sets in Table 4 are of the same order of magnitude (20-50) as we usually find when we iteratively optimize for a local density (Leydesdorff & Cozzens, 1993; Leydesdorff, 2002b).

In summary, the sciences are organized in sets of journals of an order of magnitude between 10 and 100. These journal structures represent specialties, but they are interwoven in different dimensions. For example, the set above is based on approaching the 'biochemistry' set of 97 journals from the perspective of 'chemistry.' Had we approached the set using the first dimension of the rotated solution which was spearheaded by the *Journal of Biological Chemistry,* we would have found a different delineation of the biochemistry set, that is, in relation to 'molecular biology.'

Furthermore, the ISI lists 261 journals under the category of 'Biochemistry & Molecular biology,' while our method provided us with a list of 97 journals within the chemistry set of 585 journals. If we approach the 'biochemistry' set from the decomposition on the basis of the first factor (*Journal of Biological Chemistry*), we obtain first a list of 264 journals with positive factor scores on the sixth factor, which can be designated as 'biochemistry & molecular biology,' and in a next decomposition we can further refine this set into a core set of 63 journals in the area of biochemistry. Figure 12 shows the lack of overlap among these three classifications.

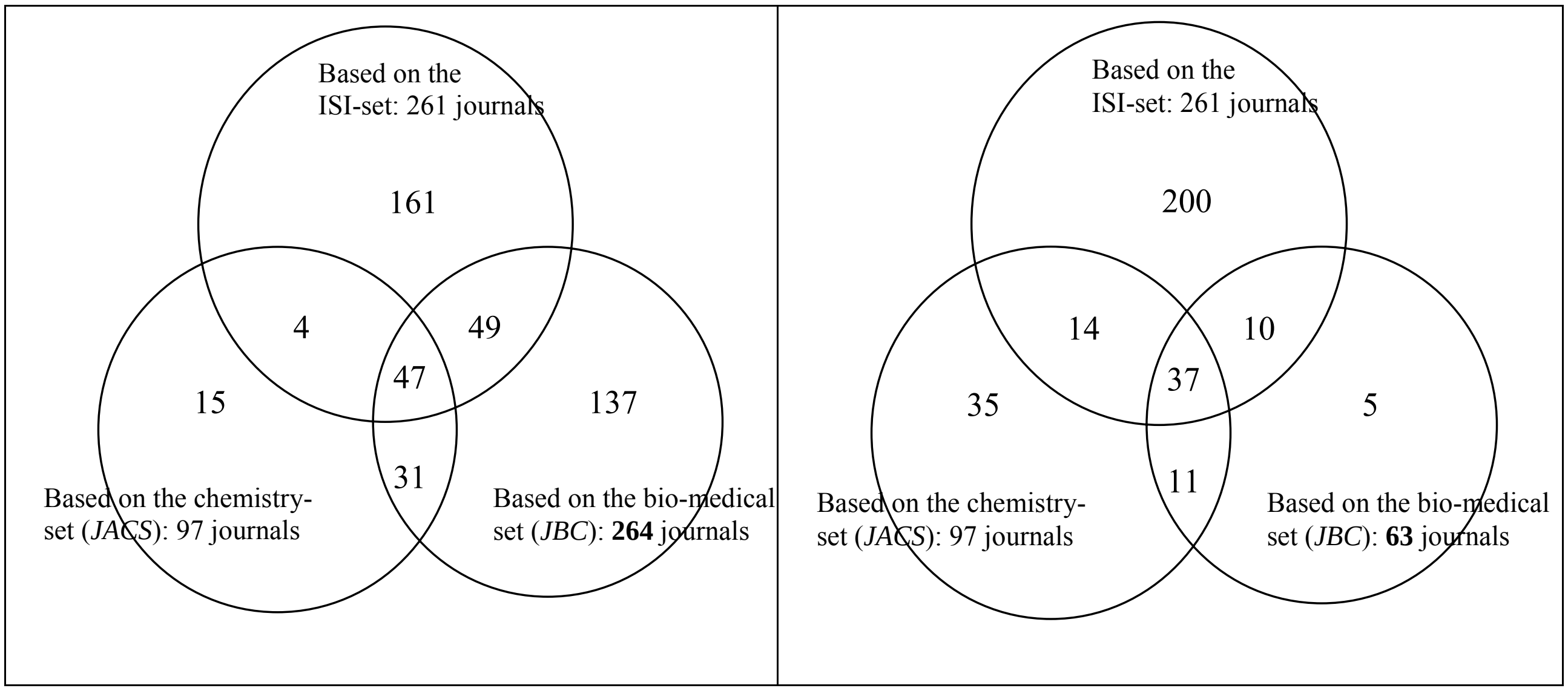

**Figure 12.** Intersections and differences among three perspectives on 'biochemistry & molecular biology' journals.

The five journals, for example, which are exclusively retrieved when using the 63 journals defined as biochemistry in the right-most circle, are *Biofizika*, the *Journal of Bioenergetics and Biomembranes*, the *Journal of General Physiology*, the *Journal of Insect Physiology*, and *Photosynthesis Research*. These five journals have negative factor-scores on the third rotated component indicating 'chemistry,' and are also not recognized in the ISI classification as belonging to the set of 'biochemistry and molecular biology.' Thus, journals can belong to a domain in different respects, and any classification uses one or more of these respects in organizing the sets.

## 5. Conclusions and discussion

The factor scores provide us with a measure of the position ('standing') of a journal in its citation environment that takes account of the size of the journal (in terms of its total citations; Bensman, 2004; Pudovkin & Garfield, 2002). In the unrotated case these scores are exact and the extraction can proceed stepwise. However, the graphs revealed the possibility of a decomposition along *diagonal* axes already in the case of the third component of the set. The analytical advantages of exact decomposition using principal component analysis, therefore, does not outweigh the clarity of the interpretation of the rotated solutions. However, the rotated factor model contains an assumption about the dimensionality of the matrix. This assumption can be theorized or pragmatically chosen (e.g., for purposes of library management). In general, the strategy of decomposing the matrix matters for the result that one obtains, because the major divisions are intersected by minor divisions like the woof and the warp of a texture (but in more than two dimensions).

The rotated factor scores enable us to focus on specific subsets with internal coherence (in terms of the covariance among the vectors of the citations). For example, the single component indicated along the diagonal of Figure 6 could be decomposed into the two rotated components of 'electronic communication' and 'microwave research' that are organized along orthogonal axes (Figure 7). Using this highly specific representation, however, the relations with the next-order level of organization—in this case the field of opto-electronics—were no longer visible. For the mapping one would need an approach which focuses on the local optima (Figure 8). Hierarchical decomposition allows for the gradual refinement of the classification in terms of disciplines and specialties. This was shown for the chemistry and then for the biochemistry factor in more detail.

The twelve rotated components at the level of the database were different from the twelve categories suggested by Glänzel & Schubert (2003), and the hundred main components after rotation were different from the 170 categories provided by the ISI at the Web-of-Science. The ISI listings are not by definition more inclusive than the analytical solutions based on rotated component analysis. For example, the ISI lists 46 journals as belonging to the group of "Inorganic and Nuclear Chemistry," while we found 106 journals under this category (Table 3). Obviously, the two measures indicate different delineations and dimensions. The overall conclusion seems to be that one cannot develop a conclusive classification on the basis of analytical arguments. However, the quality of a logical classification proposed for one reason or another can be tested against the structure in the citation database. For example, if one assumes that science is subdivided into twelve main categories, a deviation from the twelve categories provided in Table 2 suggests the need for a further reflection on the quality of this representation.